\documentclass[aip,apl,reprint,superscriptaddress,a4paper,floatfix]{revtex4-1}

\usepackage{graphicx}
\usepackage{dcolumn}
\usepackage{bm}
\usepackage{booktabs}

\begin{document}

\title{Strongly enhanced magnetic moments in ferromagnetic FeMnP$_{0.50}$Si$_{0.50}$}

 \author{Matthias Hudl} 
 \email{matthias.hudl@angstrom.uu.se} 
 \affiliation{Department of Engineering Sciences, Uppsala University, Box 534, SE-751 21 Uppsala Sweden}
 
 \author{Lennart H\"aggstr\"om}
 \affiliation{Department of Physics and Astronomy, Uppsala University, Box 516, SE-751 20 Uppsala Sweden}
 
 \author{Erna-Krisztina Delczeg-Czirjak}
 \affiliation{Department of Materials Science and Engineering, Royal Institute of Technology, SE-100 44 Stockholm Sweden}

 \author{Viktor H\"oglin}
 \affiliation{Department of Materials Chemistry, Uppsala University, Box 538, SE-751 21, Uppsala Sweden} 
 
 \author{Martin Sahlberg}
 \affiliation{Department of Materials Chemistry, Uppsala University, Box 538, SE-751 21, Uppsala Sweden}

 \author{Levente Vitos}
 \affiliation{Department of Physics and Astronomy, Uppsala University, Box 516, SE-751 20 Uppsala Sweden}
 \affiliation{Department of Materials Science and Engineering, Royal Institute of Technology, SE-100 44 Stockholm Sweden}
 
 \author{Olle Eriksson}
 \affiliation{Department of Physics and Astronomy, Uppsala University, Box 516, SE-751 20 Uppsala Sweden}  
   
 \author{Per Nordblad}
 \affiliation{Department of Engineering Sciences, Uppsala University, Box 534, SE-751 21 Uppsala Sweden}
 
 \author{Yvonne Andersson}
 \affiliation{Department of Materials Chemistry, Uppsala University, Box 538, SE-751 21, Uppsala Sweden}

\begin{abstract}


The compound FeMnP$_{0.5}$Si$_{0.5}$ has been studied by magnetic measurements, M\"ossbauer spectroscopy and electronic structure and total energy calculations. An unexpected high magnetic hyperfine field for Fe atoms located at the tetrahedral Me(1) site in the Fe$_2$P structure is found. The saturation moment derived from magnetic measurements corresponds to 4.4 $\mu_B$/f.u. at low temperatures, a value substantially higher than previously reported, but in accord with the results from our electron structure calculations. This high saturation moment, a first order nature of the ferromagnetic transition and a tunable transition temperature make the Fe$_{2-x}$Mn$_x$P$_{1-y}$Si$_y$ system promising for magnetocaloric applications.

\end{abstract}

\date{\today}


\maketitle


Fe$_2$P based compounds are promising materials for magnetocaloric applications. A study of the FeMnP$_{1-y}$Si$_{y}$ series by Cam Thanh et al. showed strong magnetocaloric effects \cite{Thanh08}. It was found that FeMnP$_{1-y}$Si$_{y}$ crystallizes in the hexagonal Fe$_2$P-type structure which persists for a Si content from y $\approx$ 0.24 up to y $\approx$ 0.65 and undergoes a first order para- to ferromagnetic phase transition with T$_C$ tunable around room temperature. The low temperature saturation moment for FeMnP$_{0.50}$Si$_{0.50}$ was of the order 3.8 $\mu_B$ per formula unit (f.u.)\cite{Thanh08}.

In the hexagonal Fe$_2$P-type and the closely related orthorhombic Co$_2$P-type (y $\leq$ 0.24) structures two equally populated metal sites are present, the tetragonally coordinated Me(1) and the pyramidally coordinated Me(2) site \cite{Rundqvist59}. The initial compound Fe$_2$P has a saturation moment of 2.94 $\mu_B$/f.u. with the site specific magnetic moments of $\mu$(Fe(1),Fe(2)) = (1.03, 1.91)$\mu_B$ and a magnetic hyperfine field of B$_{hf}$(Fe(1),Fe(2)) = (-11.4, -18.0) T, Ref.~\onlinecite{Eriksson88}. By substitution of Fe for Mn antiferromagnetic ordering and a structural phase transition into the orthorhombic Co$_2$P-type structure are induced. In pure FeMnP, the  Fe atoms preferentially populate the Me(1) site and the Mn atoms the Me(2) site \cite{Sjostrom88}. Substitution of P for Si restores the hexagonal Fe$_2$P-type structure and stabilizes ferromagnetic ordering. At the border region between the orthorhombic and hexagonal structure, y $\approx$ 0.25, the compound with hexagonal structure could either be in an antiferromagnetic or a ferrimagnetic state depending on the heat treatment after the synthesis \cite{Hudl10}. 

In this study of the  FeMnP$_{0.5}$Si$_{0.5}$ compound it is found that the magnetic hyperfine field for Fe atoms located at the tetrahedral Me(1) site is strongly enhanced in comparison to other compounds within the Fe$_{2-x}$Mn$_x$P$_{1-y}$Si$_y$ system. The enhanced magnetic hyperfine field is accompanied by a correspondingly high saturation moment of 4.4 $\mu_B$/f.u. at 5 K. An enhanced saturation moment can promote the magnetocaloric properties and make this material suitable for applications. In addition,  the FeMnP$_{1-y}$Si$_{y}$ series is favorable for applications due to its cheap, nonhazardous and environmentally friendly element composition. 
  

Samples of stoichiometric FeMnP$_{0.5}$Si$_{0.5}$ were prepared by the drop synthesis method \cite{Rundqvist59}. Fabrication details and structural characterization are reported elsewhere \cite{Hoeglin10}.
The magnetic properties were investigated by means of magnetization measurements using a vibrating sample magnetometer and a SQUID magnetometer (Quantum Design PPMS and MPMS). M\"ossbauer absorption spectra on samples of composition y = 0.46 and 0.50 were recorded in the constant acceleration mode at temperatures between 5 K and 440 K using a $^{57}$CoRh source.

The electronic structure and total energy calculations were performed using the exact muffin-tin orbital method (EMTO) \cite{Andersen94,Vitos01} in combination with the coherent potential approximation (CPA) \cite{Soven67,Vitos01a}. The ab initio calculations were carried out for three different phases of FeMnP$_{0.5}$Si$_{0.5}$. Two ordered phases were considered: one with Mn atoms occupying the pyramidal (high moment) positions and one with Mn atoms occupying the tetrahedral (low moment) positions. These structures are labeled  "Mn-pyramidal" and "Fe-pyramidal", respectively. In the third case, the Mn and Fe atoms are randomly distributed on the Me(1) and Me(2) positions, referred to as "Disordered". It was assumed that all these phases have the hexagonal Fe$_2$P structure with the elements P and Si uniformly distributed over the pnictide sites. The internal positions and lattice parameters were taken from structure refinements of neutron powder diffraction data by H\"oglin et al. \cite{Hoeglin10}. The numerical details of the calculations are similar to those reported in Ref.~\onlinecite{Hudl10}.

\begin{figure}[b]
  \centering
  \includegraphics[width=0.42\textwidth]{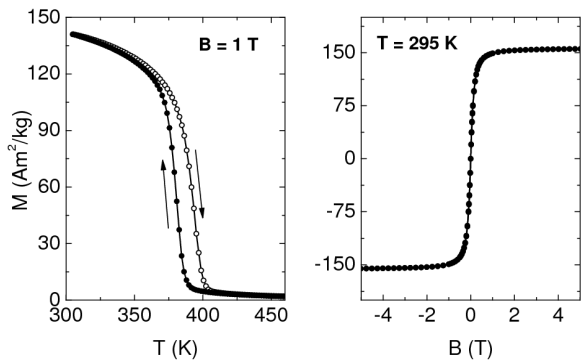}
	\caption{Magnetization data for FeMnP$_{0.5}$Si$_{0.5}$; Left graph: Magnetization vs. temperature measured in 1 T field. Right graph: Magnetization vs. magnetic field at 295 K.}
	\label{fig1}
\end{figure}

Magnetization measurements indicate a first order para- to ferromagnetic phase transition at 382 K (measured on cooling) as shown in Fig. \ref{fig1}. The magnetic phase transition is accompanied by a 7-12 K wide thermal hysteresis region. In the magnetization vs. magnetic field data a soft magnetic behavior with marginal magnetic hysteresis is seen. At 295 K the saturation magnetization is 156 Am$^2$/kg which correspond to 3.9 $\mu_B$/f.u., see Fig. \ref{fig1}. At 5 K the saturation moment is 4.4 $\mu_B$/f.u. which is higher than found in previous experimental studies, 3.8 $\mu_B$/f.u., \cite{Thanh08}, and slightly higher than earlier theoretical calculations indicate, 4.2 $\mu_B$/f.u., \cite{Divis08}. The magnetic entropy change was estimated from magnetization data and for a magnetic field change of 1.8 T  a magnetic entropy change of about 8 J/kgK is obtained.

The room temperature XRD pattern for FeMnP$_{0.5}$Si$_{0.5}$ reveals a hexagonal Fe$_2$P-type structure, space group P-62m, with lattice parameters \textit{a}=6.2090(3) \AA{ }and \textit{c}=3.2880(3) \AA{}. The first order magnetic transition is accompanied by a structural transition where the lattice parameter \textit{a} decreases by 2 \%, \textit{c} increases by about 5 \% and the cell volume increases approx. 1 \%,  however the compound remains in the Fe$_2$P-type structure\cite{Hoeglin10}.


\begin{figure}[htb]
  \centering
  \includegraphics[width=0.30\textwidth]{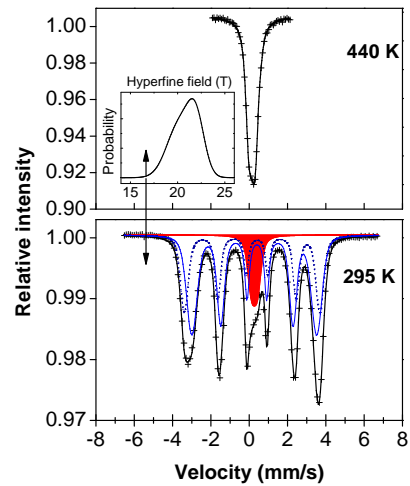}
	\caption{M\"ossbauer spectra for FeMnP$_{0.5}$Si$_{0.5}$ measured at 295 and 440 K. Inset: Magnetic hyperfine field distribution at 295 K.}
	\label{fig2}
\end{figure} 

M\"ossbauer spectra for y $=$ 0.50 at 440 K and at 295 K are shown in Fig. \ref{fig2}. The spectra in the paramagnetic regime show a broad single line centered at around 0.17 mm/s. The average electric quadrupole splitting was  0.32 mm/s. Reducing the isomer shift value with the high temperature limit of -7.3 $\cdot$ 10$^{-4}$ $\Delta$T for the second order Doppler shift gives an isomer shift value $\delta$ = 0.28 mm/s at room temperature, which is in good agreement with room temperature values for Fe(1) in hexagonal Fe$_{2}$P$_{1-y}$Si$_{y}$, \cite{Srivastava87} and in FeMnP$_{0.75}$Si$_{0.25}$ \cite{Hudl10}. This indicates that Fe only populates Me(1) in the present samples since the isomer shift values at room temperature for Me(2) in these compounds are much larger, being  0.54 mm/s (Fe$_2$P) and 0.57 mm/s (FeMnP$_{0.75}$Si$_{0.25}$).
 
\begin{figure}[ht]
  \centering
  \includegraphics[width=0.32\textwidth]{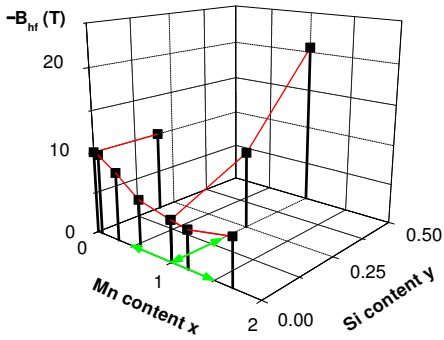}
	\caption{Magnetic hyperfine fields B$_{hf}$  for Fe at the tetrahedral Me(1) site for different Fe$_{2-x}$Mn$_x$P$_{1-y}$Si$_y$ compounds. All compounds show hexagonal crystal structure except along the green lines around FeMnP which have the  closely related orthorhombic Co$_2$P structure. Data taken from this study and references \cite{Hudl10,Sjostrom88,Srivastava87,Jernberg84}.}
	\label{fig3}
\end{figure}

The spectra in the ferromagnetic regime show a well resolved six-line pattern together with a central broad line. Due to the difference in the line intensities between outer lines the spectra were fitted with two sextets and an electric quadrupole splitted doublet. The doublet was absent at 77 K and below. The average hyperfine interaction parameters at 295 K for the Fe(1) sextet were: the magnetic hyperfine field B$_{hf}$ = 20.6(3) T and the isomer shift $\delta$ = 0.30(1) mm/s. At 5 K, B$_{hf}$ was found to be 22.8(3) T. Figure \ref{fig3} shows the magnetic hyperfine fields for different Fe$_{2-x}$Mn$_x$P$_{1-y}$Si$_y$ compounds from the current investigation and earlier publications\cite{Hudl10,Sjostrom88,Srivastava87,Jernberg84}. 
The relation between the magnetic hyperfine field and the magnetic moment is not straightforward but can be written as B$_{hf}$ = -12.6$\mu$ + B$_v$ (omitting small dipolar and orbital contributions)\cite{Eriksson88}. For the present compound the s-valence term was calculated to B$_v$(Me(1), Me(2)) = (-1.96, 17.1) T. Using the experimental B$_{hf}$ we arrive at a saturation Fe(1) moment of 1.65 $\mu_B$ which corresponds to an increase of 60\% compared with 1.03 $\mu_B$ in Fe$_2$P .
For the closely related compound Fe$_2$P$_{1-y}$Si$_y$ an almost linear relation between the magnetic moments and the average nearest neighbor (nn) Fe-P distance is known. \cite{Severin95} The average nn Fe(1) to P distance in Fe$_2$P is 2.26 \AA { } while the corresponding distance in FeMnP$_{0.5}$Si$_{0.5}$ is 2.32 \AA. Using the linear relation with the found lattice expansion would correspond to an increase of the Fe(1) moment from 1.03 $\mu_B$ to 1.24 $\mu_B$. 
Additionally, first-principle band calculations on the hexagonal compound FeMnP$_{0.66}$Si$_{0.33}$ \cite{Liu09} showed an Fe(1) moment of 1.35 $\mu_B$ for random occupation of Si atoms over the pnictide sites but an increase to 1.46 $\mu_B$ for Si preferentially occupying the P(1) site.  For both, lattice expansion and Si ordering on the pnictide sites an increase of the magnetic moment is known. 

\begin{figure}[ht]
  \centering
  \includegraphics[width=0.28\textwidth]{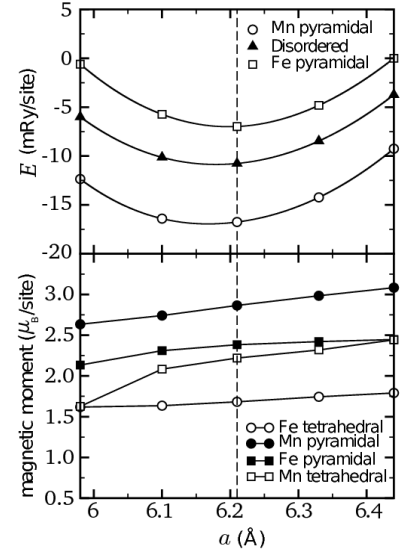}
	\caption{Theoretical results for FeMnP$_{0.5}$Si$_{0.5}$. Total energy per site (upper panel) and site-projected magnetic moments (lower panel) of the Mn-pyramidal- (cirles) and Fe-pyramidal (squares) Ordered phases as a function of the lattice parameter \textit{a}  and for fixed \textit{c/a} $=$ 0.5296. The dashed line indicates the experimental lattice parameter.}
	\label{fig4}
\end{figure}

Calculated total energies as a function of the lattice parameter \textit{a} (for fixed \textit{c/a} $=$ 0.5296) for three different phases (defined above): Mn-pyramidal, Fe-pyramidal and Disordered are shown in Figure \ref{fig4} (upper panel). We find that the Mn-pyramidal phase has the lowest energy for all volumes (lattice parameters). This is in line with the experimental observation, namely that Mn atoms preferentially occupy the high moment site. The energy difference between the considered phases is rather significant (5-10 mRy/site). The Mn pyramidal phase has a shallow energy minimum for lattice constants around 6.16 \AA, whereas our experimental value is 6.209 \AA{ }for the hexagonal FeMnP$_{0.5}$Si$_{0.5}$.  The site projected magnetic moments of the Fe and Mn atoms for the ordered phases are displayed in Fig. \ref{fig4} (lower panel). Note that the magnetic moments of the Fe and Mn atoms are always ferromagnetically aligned in the ordered phases. A decrease of the magnetic moments towards lower volumes occurs due to the magnetovolume effect\cite{Mohn06}. Energetically the Mn-pyramidal phase is the stable one which corresponds to Mn atoms occupying the high moment site, with Mn(Fe) moment 2.81(1.64) $\mu_B$/atom for the equilibrium volume (lattice parameter). In the Fe-pyramidal phase the Mn moments are always lower than in the Mn-pyramidal phase, and it is tempting to explain the stabilization of the Mn-pyramidal phase to be due to the exchange energy of the larger Mn moment in the pyramidal site. These calculations give a total magnetic moment of 4.3 $\mu_B$  per formula unit, which is in good agreement with our experimental measurements and with previous theory \cite{Divis08}. 


From M\"ossbauer spectroscopy it is found that Fe atoms in FeMnP$_{0.5}$Si$_{0.5}$ occupy the tetrahedral Me(1) site. Our ab initio caculations support this distribution of the metal atoms and indicate a ferromagnetic structure with a total moment of 4.3 $\mu_B$, in good agreement with the measured low temperature saturation moment, 4.4 $\mu_B$ per formula unit. The low temperature magnetic hyperfine field, -22.8 T, for Fe at the Me(1) site is much larger than earlier results for compounds in the Fe$_{2-x}$Mn$_{x}$P$_{1-y}$Si$_{y}$ system, cf. Fig. \ref{fig3}. However, this value of the hyperfine field  indicates a magnetic moment of 1.65 $\mu_B$ for Fe(1), which agrees  well with the calculated value  1.64 $\mu_B$. This large magnetic moment is especially interesting since it can enhance the magnetocaloric properties of the  Fe$_{2-x}$Mn$_{x}$P$_{1-y}$Si$_{y}$ system. 

Financial support from the Swedish Energy Agency (STEM) and the Swedish Research Council (VR) is acknowledged. Supercomputer resources provide by SNIC are acknowledged.



\end{document}